# Fish-tail effect and irreversibility field of

# (Cu,C)Ba$_2$Ca$_3$Cu$_4$O$_x$-(LiF)$_y$ superconductor

[1,2,*]P. Badica, [2]V. Sandu, [2]G. Aldica, [1,3]A. Iyo, [1,3]H. Kito, [1,4]M. Hirai and [1,3]Y. Tanaka

[1] Nanoelectronics Research Institute of National Institute for Advanced Industrial Science and Technology, 1-1-1 Umezono, Tsukuba, Ibaraki, 305-8568 Japan

[2] National Institute for Materials Physics, POBox MG-7, RO-76900, Romania

[3] CREST JST, Kawaguchi, Saitama, 332-0012 Japan

[4] Tokyo University of Science, 2641 Yamazaki, Noda, Chiba 278-8510 Japan

contact author: P. Badica

[*] present address: Institute for Materials Research, Tohoku University, 2-1-1 Katahira, Aoba-ku, Sendai, 980-8577, Japan

e-mail: p.badica@imr.tohoku.ac.jp

fax: 81-22-215-2091

# Fish-tail effect and irreversibility field of (Cu,C)Ba$_2$Ca$_3$Cu$_4$O$_X$-(LiF)$_y$ superconductor

[1,2,*]P. Badica, [2]V. Sandu, [2]G. Aldica, [1,3]A. Iyo, [1,3]H. Kito, [1,4]M. Hirai and [1,3]Y. Tanaka

[1] Nanoelectronics Research Institute of National Institute for Advanced Industrial Science and Technology, 1-1-1 Umezono, Tsukuba, Ibaraki, 305-8568 Japan

[2] National Institute for Materials Physics, POBox MG-7, RO-76900, Romania

[3] CREST JST, Kawaguchi, Saitama, 332-0012 Japan

[4] Tokyo University of Science, 2641 Yamazaki, Noda, Chiba 278-8510, Japan

**Abstract**

Addition of (LiF)$_{y \leq 0.15}$, and of proper amount of (AgO)$_{z=0.45-0.8}$ as oxidizing agent, to (Cu,C)Ba$_2$Ca$_3$Cu$_4$O$_{10+\delta}$ superconductor is useful to control and to shift the doping characteristics (hole density and distribution, and level of disorder) into the region where the irreversible properties, i.e. fish-tail effect (FTE) and irreversibility field $H_{irr}$ are improved. Among notable effects are the development of the second magnetization peak with a higher amplitude $J_{c,\,max}$ and the enhancement of $H_{irr}$ at high temperatures, above a certain value T$^*$ which depends on both y$_{LiF}$ and $z_{AgO}$. The best results are obtained for the sample with $y_{LiF}$=0.1 and $z_{AgO}$=0.73. This sample preserves its single phase Cu,C-1234 composition. The influence on the FTE and H$_{irr}$ of the interplay between doping characteristics, controlled by LiF and AgO content, is discussed.

Keywords: (Cu,C)Ba$_2$Ca$_3$Cu$_4$O$_{10\pm\delta}$, LiF-addition, fish-tail effect, irreversibility field, high-T$_c$ superconductor

# 1. INTRODUCTION

Magnetic flux line system in high temperature superconductors (HTS) is a prolific world of interesting phenomena that strongly enriched and changed our representation on many physical concepts. One of the most important challenges for the HTS classes with critical temperatures $T_C$ higher than 100 K is the enhancement of the intra-granular critical current density $J_C$, i.e. the irreversible magnetization $\Delta M$, at high fields and elevated temperatures ($T > 77$ K). Here, $\Delta M = \dfrac{M^+ - M^-}{2}$, where $M^\pm$ stands for magnetization on the ascending (+) and descending (-) branches of the hysteretic loop $M$ vs $H$. Associated with the problem of irreversibility and of special interest for the improvement of the transport properties of HTS is anomalous increase of $J_C$ in a certain field range, also known as the *fishtail effect* (FTE). The effect was extensively studied in the main classes of HTS and it was established that, except for $Bi_2Sr_2CaCu_2O_8$, the amplitude of the peak is rather small, but extends on a considerable field range. FTE is supposed to be a succession of crossovers and phase transitions in the vortex line system triggered by field that permits a better matching of the flux line system to the pinning centers. The effect is different from the second peak effect observed in low $T_c$ superconductors and very pure $CuBa_2YCu_2O_{7-\delta}$ [1] single crystals, close to $H_{c2}$. Specifically, the latter is the result of the fast softening of the shear module $C_{66}$ of the flux line lattice [2] close to $H_{c2}$, whereas FTE is the result of a more subtle interplay between pinning, thermal, and elastic energies of the vortex lattice [3-5].

$(Cu,C)Ba_2Ca_3Cu_4O_{10+\delta}$ compound, thereafter (Cu,C-1234), despite a very high critical temperature, has rather low and rapid decreasing with $T$ and $H$ critical current density. Therefore, it is of great interest to find a way to improve this material. It is not a trivial problem to obtain both an optimal doping with holes and enough strong pinning centers due to the special way of preparation of this material (high pressure synthesis). A finer triggering of the transport properties is by the addition of a "hole controller" apart from the oxidizing agent (AgO). For this aim, we chose lithium inserted via LiF. A detailed investigation of the effect of lithium and its interplay with oxygenation on the critical current density $J_C$ and irreversibility field $H_{irr}$ is the main goal of the present paper. The subject is of interest also considering that just few articles report about FTE in this superconductor [6, 7]. The choice of lithium relies upon its *friendly* behavior noticed in the previous experiments with the bilayered $CuBa_2YCu_2O_{7-\delta}$ [8-10].

Cu,C-1234 has four $CuO_2$ layers sandwiching calcium stacked along the *c*-axis, separated by a blocking BaO layer and a $CuO_{1-\delta}$ charge reservoir [11-14]. In this complex structure the charge doping depends on either the oxygen control or cation substitution [15, 16]. The charge doping is of capital importance in HTS because it is accepted that changes in the charge carrier density or its distribution brings on severe modifications in the electrical and magnetic properties, including the occurrence of FTE.

There are no available data concerning the site preferred by lithium and fluorine. Due to its comparable ionic radius, lithium ($r_{Li^+} = 0.76\,\text{Å}$) most likely substitutes copper ions ($r_{Cu^{2+}} = 0.73\,\text{Å}$), but the presence in some samples of the $Ca_{0.8}CuO_2$ impurity suggests that calcium site could be also chosen ($r_{Ca^+} = 0.99\,\text{Å}$). Concerning fluorine location, it it is supposed to substitute the apical oxygen due to their close ionic radii, $r_{F^-} = 1.19\,\text{Å}$ and $r_{O^{2-}} = 1.28\,\text{Å}$. This difference of the ionic radii produces some local shrinkage of the blocking blocks, which are able to depress the superconducting order parameter. However, despite the lack of precise information, our previous experiments [17] indicate that in the LiF-doped Cu,C-1234 samples, the donor effect of Li is compensated by the oxygen supplied from an AgO-source, which also wash out the effect of fluorine. We have also observed that for a given amount of LiF, deviation from the optimum concentration of oxygen, controlled through the addition of AgO, can produce biphasic samples [17]. Although our investigations will focus mainly on the single phase Cu,C-1234-LiF sample in comparison with the un-doped compound, some results on the biphasic systems will be also presented for a better understanding of the phenomena.

## 2. EXPERIMENTAL

Samples with the starting composition $(Cu_{0.6}C_{0.4})Ba_2Ca_3Cu_4O_x$-$(LiF)_{y\leq0.15}$ were prepared by high pressure synthesis method from a $Ba_2Ca_{2.7}Cu_{4.6}C_kO_x$ precursor, $CaCO_3$, LiF and AgO as oxidizing agent. The precursor powder was obtained by solid-state route (890°C / 24 h /$O_2$-flow) from barium and calcium carbonates, and CuO. The carbon content in the precursor powder, as determined by Fourier Transformed Infrared Spectroscopy (FT-IR) [18], was $k = 0.1$. Mixtures of powders were sealed in Au-capsules and treated at 980°C for 2 h under a pressure of 3 GPa. The oxidizing agent AgO was varied between $z_{AgO}$ = 0.45 and 0.80 mol and the molar ratio *precursor/CaCO3/AgO* was 1/0.3/$z$. The samples were named as *PN*, $N$ = 1, 2, 3, 4, 5, and 6 depending on the starting amount of AgO and LiF. The sample labeled *P0* is single phase, LiF-free Cu,C-1234 and has been obtained by the same procedure for $z_{AgO}$ = 0.45. Samples labeled *P1, P2,* and *P3* are single phase Cu,C-1234-LiF. Samples *P4* and *P5* are biphasic Cu,C-(1234)-LiF and 1245)-LiF, whereas sample *P6* is a mixture of Cu,C-1223-LiF and Cu,C-1234-LiF. Samples *P2, P4, P5,* and *P6* were prepared with the same content of LiF, $y_{LiF}$ = 0.1. The composition of each sample is given in the Table 1 together with other characteristics.

As-prepared bulk samples were characterized by X-ray diffraction, scanning electron microscopy (SEM), resistance $R$ vs. temperature $T$ ($R(T)$) and thermopower (Seebeck coefficient, $S_{293K}$) measurements [17]. Noteworthy, $R(T)$ dependence of all samples exhibits an outstanding linearity up to the room temperature with a negative residual resistivity, $\rho = \rho_0 + aT$, $\rho_0 < 0$. However, a small convexity is detected for sample *P4* above $T_c$ suggesting a small hole underdoping, while the sample *P2* shows a slight concavity as for overdoped HTS. The normal-superconducting transition, with the critical temperature $T_c$, in the range 111 – 115 K (Table 1) is very narrow showing a maximum width of $\Delta T_c \approx 1$ K. The Seebeck coefficient $S_{293K}$ of the samples *P0, P1, P2, P3* and *P6* is negative, whereas it is close to zero for *P4* and *P5* samples (Table 1).

DC magnetization loops were measured with a SQUID magnetometer (MPMS Quantum Design) up to an applied field of 7 T for 20 K≤ $T$ ≤ 102 K.

## 3. RESULTS and DISCUSSION

All hysteresis loops exhibit sharp low field peaks and a counterclockwise rotation that is more emphasized in Li-doped sample (Fig.1). The sharpness of the first peak increases with $T$ indicating a good homogeneity of the sample, but with low bulk pinning. The rotation is typical for superconducting samples with paramagnetic ions. For $H >> H_{irr}$, the equilibrium magnetization can be approximated as $M_{eq} = \dfrac{M^+ + M^-}{2} = \chi(T)H - f(H)H_{c1}(T)$ and shows a rapid increase when LiF is added. However, the susceptibility of the single-phase samples $P0, P1, P2, P3$ at 77 K is approximately constant. The increase of paramagnetism could arise mainly from Cu $3d$ electrons, but these electrons are expected to participate in superconductivity. Therefore, the enhanced paramagnetism signalizes an increased amount of nanodomains with a reduced concentration of holes, most likely located close to the lithium sites. To compensate this localization of holes around lithium, addition of more oxygen is required. As can be seen from Fig 2 the increase of AgO content decreases the paramagnetism to the same level as in the LiF-free sample. However, the susceptibility of the single-phase sample $P2$ at 77 K is with 14% higher than in LiF-free $P0$ sample.

The critical current density $J_C$ was obtained using the Bean relationship for fields higher than the full penetration fields, $J_c \propto \dfrac{\Delta M}{D}$, with $D$ the average grain size. In order to avoid rough approximation due to the spread of $D$ within sample, we used the ratio $J = \dfrac{J_c}{J_{c,p}}$, where $J_{c,p}$ is the current density at the first magnetization peak.

## 3.1 Fish-tail effect

All samples investigated in this work show FTE except *P1*; for example, in Fig.3 are presented the relative critical current densities $J=J_c/J_{c,p}$ for the (Cu, C)-1234 single phase samples *P0* (LiF-free, Fig. 3a) and *P2* (LiF-added Fig. 3b), at temperatures between 20 and 102 K, while in the Fig. 3c are gathered *J-H* curves at 77K for the samples exhibiting FTE. Because FTE has a weak amplitude as in previously reported data on Cu-1234 [12, 13], we subtracted the "*background*" from the *J(H)* curves for a better characterization. The *background* $J_b$ was obtained by fitting the beginning and the tail of the experimental curve *J*, outside the peak range, with an appropriate combination of polynomial and exponential functions. The procedure allowed us to obtain the shape and field dependence of the fish-tail-current density, $J_{FT} = J - J_b$, and results at 77K are presented in Fig. 4 and for the sample with the strongest FTE (at different temperatures), *P2*, are shown in Fig. 5.

Sample *P0* without addition of LiF will be considered as optimum doped ($S_{293K}$ = -3.6μVK$^{-1}$) because it reaches the maximum $T_c$. For this sample, FTE is almost indistinguishable at low temperatures and only at 60 and 77 K is barely discernible (follow the arrows in Fig. 3 a). No FTE was observed in the slightly overdoped ($S_{293K}$ = -5μVK$^{-1}$) sample *P1* with low content of LiF ($y_{LiF}$ = 0.05) and AgO ($z_{AgO}$ = 0.59). The other two samples *P4* and *P5,* also with low amount of AgO and fixed $y_{LiF}$=0.1, which are underdoped ($S_{293K} \approx 0$ μVK$^{-1}$) show a weak FTE (Fig. 3 and Fig. 4), but stronger than for *P0*. On the other side, all samples with high $z_{AgO}$ and $y_{LiF}$ = 0.1-0.15, *P2, P3 and P6*, exhibit relatively strong FTE at 77K; the maximum amplitude of the FTE-peak, $J_{FT,max}$ (Fig. 4 inset), is attained for the sample *P2* ($y_{LiF}$ = 0.1 and $z_{AgO}$ = 0.73) and the value is more than 2 time higher than for *P4* and *P5*. At low fields (compare Fig 3a and Fig. 3b for a certain temperature), the critical current density *J* of *P2* is nearly twice lower than in *P0* but, *nota bene*, decreases much slower with increasing field and exhibits FTE above a certain onset field (Fig.3b). This onset field shifts to lower field as the temperature is increased (Fig. 5). This sample has the strongest overdoping ($S_{293K}$ =-8 $\mu VK^{-1}$) and, remarkably, it is single phase Cu,C-1234. Not only for *P2*, but also for the other single phase sample *P3*, at constant temperature, the field dependence of $J_{FT}$ is well fitted with the log-normal function, $J_{FT} = A\exp(-0.5(\ln(H/Hp)/\sigma)^2)$. The width of $J_{FT}$, i.e. the field range where the amplitude is suppressed *e* times, decreases linearly when *T* increases (see Fig.5 for the sample *P2* where FTE can be easily observed, i.e. between 40 and 102 K). Hagen and Griessen [19] obtained similar dependence for the distribution function of the flux activation energy in polycrystalline $CuBa_2YCu_2O_{7-\delta}$. This indicates that the extended FTE in this sample is a succession of crossovers and transitions controlled by the magnetic field.

It is worth mentioning here that the other samples are biphasic: samples *P4 and P5* are composed of Cu,C-1245 and Cu,C-1234, while *P6* contains Cu,C-1234 and Cu,C-1223 (Table 1). We would like also to note that in fact FTE is composed of two peaks. Samples *P4* and *P5* with low $z_{AgO}$ show a stronger first FTE-peak at low fields (Fig. 4), while samples with high $z_{AgO}$, *P2*, *P3*, and *P6* develop a second FTE-peak at higher fields. However, in the samples *P4*, *P5* ($y_{LiF}$ = 0.1 and low $z_{AgO}$) the evolution of the $J_{FT}$ with the starting amount of AgO probably reflects the evolution of the two-phase composition. Namely, below the full penetration field, the first magnetization peak is replaced by a two-step structure (see the inset to the Fig. 3c inset), like a superposition of two different peaks that belong to different phases. At high fields, these samples exhibit also a nearly imperceptible two-peak structure. The first peak, at low field ($\mu_0 H_{p1} \approx 0.35$ T) is practically $z_{AgO}$-independent and is narrow, but six times stronger than the second peak at high fields ($\mu_0 H_{p2} \geq 3.5$ T, Fig 4).

Our data suggest that an increase of $z_{AgO}$ enhances the amplitude of the FTE-peak and also broadens it. Although oxygen seems to have the strongest influence on FTE there is a fine interplay between the effects induced by LiF addition and those by oxygenation, as it will be revealed in the next paragraphs.

For a further insight, we proceed to a detailed analysis of the FTE and irreversibility field in LiF-added Cu,C-1234 compound. We mainly concentrate on the (almost) single-phase samples *P2* and *P3*. The peak field $H_p$ (taken as the field where $J_{FT}$ is maximum for a given temperature) shifts to lower values with increasing $T$ and for $T < 93$ K, it obeys the following experimental law as can be seen from the Fig.6:

$$\mu_0 H_p = 13.85(1 - T/T_c)^{\alpha} \text{ (T)}, \qquad (1)$$

with $\alpha$ (*P2*) ≈ 1.74 and $\alpha$ (*P3*) ≈ 1.67. Theoretical models propose a similar dependence for $H_p$, but with smaller exponent, i.e. $\alpha = 3/2$ [5] for the transition to an amorphous solid, or $\alpha = 4/3$ when above $H_p$ the system is in a strongly pinned liquid state (plastic flow) [20]. From this dependence we can get information regarding the nature of the disorder involved in the pinning process. If at $H_p$ the plastic energy barrier $E_{el} = \varepsilon \varepsilon_0 a_0$ balances the pinning energy barrier, $E_{pin} = (\gamma \varepsilon^2 \varepsilon_0 \xi^4)^{1/3} \left(\dfrac{a_0}{L_c}\right)^{1/5}$ [20], the temperature dependence of the peak field is:

$$\mu_0 H_p \propto \gamma^{-1} \varepsilon \lambda^{-4} \xi^{-3} = \gamma^{-1} \varepsilon \lambda^{-4}(0) \xi^{-3}(0)(1 - T/T_c)^{7/2} \qquad (2)$$

where $\gamma$ is the disorder strength, $\varepsilon$ the anisotropy parameter, $\xi$ and $\lambda$ the Ginzburg-Landau (GL) coherence length and penetration length, respectively, $a_0 \propto \sqrt{\dfrac{\Phi_0}{H}}$, $\varepsilon_0 = \left(\dfrac{\Phi_0}{4\pi\lambda}\right)^2$, and the Larkin length $L_c = \left(\dfrac{\xi^2 \varepsilon^4 \varepsilon_0^2}{\gamma}\right)^{1/3}$. In the Eq. 2 we assumed that $\xi$ and $\lambda$ have the universal GL dependence $\propto (1 - T/T_c)^{-1/2}$. With the experimental dependence of $H_p$ for our Cu,C-1234-LiF sample *P2* and *P3* it results that $\gamma \propto (1 - T/T_c)^{\beta}$ and $\beta$(*P2*) ≈ 1.76 and $\beta$(*P3*) ≈ 1.83, a value which is close to $\beta = 2$ predicted for $\Delta T_c$-pinning [21].

A consequence of the Eq. 2 is that the peak field shifts to higher fields when the disorder strength γ decreases. This effect could explain the evolution of $J_{FT}$ with doping shown in Fig. 4. The more disordered samples, *P4* and *P5*, are grouped around a smaller peak field ($\mu_0 H$=0.35 T) whereas the more ordered samples, *P0, P2, P3* and *P6,* are grouped between 1.5 and 1.9 T. In the latter, the disorder is at the microscopic level and seems to be predominantly due to the oxygen distribution within blocking blocks. In the *P4* and *P5* samples, due to the presence of two superconducting phases with a different number of $CuO_2$ layers, it is possible that the scale disorder is different in (Cu,C)-1245 and in (Cu,C)-1234. This difference might explain the two peaks observed in $J_{FT}$.

The amplitude of $J_{FT}$, has a strong dependence vs. $z_{AgO}$ (see the inset to Fig.4). Correlated with the resistance R(T) and thermopower measurements [17], this dependence suggests that $J_{FT,max}$ increases with the deviation $\delta p$ of the hole density from the optimally doping of $CuO_2$ layers. So, $J_{FT,max}$ has the smallest value in the quasioptimally hole-doped samples *P0, P1* and higher values either for small $z_{AgO}$ in the underdoped samples *P4* and *P5* or for high $z_{AgO}$ in the overdoped samples *P2, P3*, and *P6* (Fig.4). R vs. T curves smoothly changes from slightly convex to slightly concave with increasing $z_{AgO}$, and the Seebeck coefficient $S_{293K}$ is practically zero in the underdoped samples *P4* and *P5*, and is negative in *P0, P2*, and *P6* samples. All these agree with a continuous increase of the holes density with increasing $z_{AgO}$ and the absolute value of $S_{293K}$ is two times higher in *P2* than in *P0*. However, there is an asymmetry between the hole-overdoped and hole-underdoped samples. Roughly, this is the result of the quadratic dependence of the critical current on the holes density, $J_C \cong \frac{c}{\Phi_0}\left(\frac{\gamma}{L_C}\right)^{1/2} \propto \lambda^{-2} \propto p^2$ for $\Delta T_C$-pinning [21]. The fact that *P6* has a slightly lower $J_{FT,max}$ is possibly due to an alteration of the ratio between the density of holes from the superconducting blocks and charge reservoir which appears above a certain nominal hole concentration [22], i.e. more holes does not mean necessary more holes in the $CuO_2$ planes.

*3.2 Irreversibility field*

Irreversibility field $H_{irr}$ was obtained from the $J_c$ vs. $H$ loops using the criterion $J_c(H_{irr})$ = 1000 A·cm$^{-2}$ (grain size is estimated from SEM observations at 7-20μm for all samples, and is taken 15μm for calculations of $J_c$). The analysis for *P0, P2 and P3* samples shows that $H_{irr}$ obeys the law $H_{irr} = H_{irr}(0)\left(1 - \frac{T}{Tc}\right)^{\zeta}$ (Fig.7), with both $H_{irr,0}$ and the exponent $\zeta$ decreasing with increasing the amount of LiF i.e. $\zeta(P0)$ = 3.4, $\zeta(P2)$ = 2.9, $\zeta(P3)$ = 2.4 and $H_{irr,0}(P0)$ = 235 T, $H_{irr,0}(P2)$ = 197 T, $H_{irr,0}(P3)$ = 71 T. Subsequently, there is a certain temperature $T^*$ beyond which $H_{irr}$ is higher in the sample with more LiF content than in the samples with lower LiF addition, including Li-free sample *P0*. For example, at $T^* \geq 0.5T_C$, $H_{irr}$ is higher in the sample *P2* than in the sample *P0*. *This is the main achievement of the addition of LiF in* (Cu,C)-1234. Noteworthy, such unusual high values of $\zeta$ have also been reported in other high-pressure compounds [23].

The irreversibility line is the result of a complex interplay between the elastic $E_{el}$, pinning $E_{pin}$ and thermal $k_BT$ energies. When the thermal energy prevails over one $E_{el}$ or $E_{pin}$ the system behaves as reversible. Therefore, in a simple picture the character of the irreversibility line is given by the relative strength of $E_{el}$ and $E_{pin}$ and might be either a melting line or a depinning line. Complications could appear when the surface barriers start to be involved. It seems less probable that the addition of lithium should produce major effects on the elastic energy; but both lithium and fluorine creates local structural distortions. These distortions, even small, are centers for accumulation of oxygen vacancies, and some experimental results point towards a kind of internal granularity associated with the distribution of oxygen vacancies that creates disorder and frustration [24, 25]. Due to the small coherence length, the order parameter is extremely sensitive to the non-homogenous domains created by the clusters of vacancies. Hence, it should be expected the increase of the irreversibility. However, the amount of oxygen is crucial in this process. At low oxygen content, there are large clusters of vacancies, which constitute the weak points where the field penetrates the grains reducing the connectivity and critical currents. An increase of the oxygen content reduces the global density of vacancies, hence, diminishes the cluster size, homogenizes their distribution, and increases the number of the close metastable states pushing the limits of nonergodicity (irreversibility) to higher fields. A further increase of oxygen content by increasing $z_{AgO}$ starts to reduce the point defects, the main ingredient of the collective pinning. Indeed, this evolution of irreversibility with increasing the amount of AgO is shown in the inset to Fig. 7: the enhancement of $H_{irr}$ up to $z_{AgO} = 0.73$ is followed by a slight decrease of $H_{irr}$ for higher $z_{AgO}$.

As have been previously shown (see Table 1 and ref. 17), the addition of lithium has severe effects on the phase homogeneity of (Cu,C)-1234 compound, promoting the formation of highly layered (Cu,C)-1245 phase. The increase of the oxidizing factor AgO has a reverse effect inhibiting the formation of this phase, and if high enough favors the low layered (Cu,C)-1223. However, there is a certain AgO concentration, $z_{AgO} = 0.73$, for which the compound is single (Cu,C)-1234-(LiF)$_{0.1}$ phase. In this phase, the effect of addition (e.g. LiF) is beneficial for the intra-grain transport properties mainly for $T > T_c/2$ and this improvement has two aspects: first the FTE is sensitively increased at moderate fields and, second, the irreversibility field is increased at high temperature. The mechanisms by which such enhancement occurs are not completely understood, and several data seem to suggest irreconcilable explanations. For example, the increased paramagnetic moment points toward a local depletion (localization) of holes around a (reduced) fraction of $Cu^{3+}$ ions, whereas the peak field dependence on the oxidizing factor and $z_{AgO}$ dependence of the irreversibility field pleads for a higher level of homogenization in that samples. Moreover, considering the dependence of the critical current density on the charge density, the increase of the critical current density it is also in contradiction with the localization. It is possible that the increased number of blocks should play a more active role in the distribution of the holes over the whole structure exceeding the simple picture with holes only in the $CuO_2$ planes.

*3.3 Influence of grinding on magnetization*

The field dependence of ΔM at 77 K before and after grinding is shown in the Fig. 8 for the samples *P0* and *P2*. Here, we talk about ΔM instead of $J_c$ because it is rather difficult to use Bean's relationship to define the critical current density in the ground samples. Close to the full penetration field, ΔM decreases 2.5 times for the sample *P0* and almost 3 times for the sample *P2*. The weak FTE observed in the *P0* sample practically vanishes, while in the sample *P2* FTE decreases five times. Beside the size reduction, the mechanical process is likely to induce a certain structure containing dislocations and cracks. It is expected that the surface will play a more important role when the grain size is reduced; this is mainly due to the vortex-mirror image interaction that pushes the flux line off the grain and increases the relaxation [26, 27]. If so, the entry magnetization $M_{en}$, i.e. the magnetization for fields just above the penetration field, should obey the law $M_{en} \approx H^{-1}$ [27]. In our case, we have found a power law, $M_{en} \approx H^{-\nu}$, with ν decreasing from 0.7 to 0.56 and 0.51 for samples *P0* and *P2*, respectively, when *T* increases (Fig.9). This hints at a reduced influence of the surface on the pinning. The most plausible explanation is the size-induced crossover from the bulk pinning to a two dimensional pinning [28]. When the grain size decreases below the pinning correlation length $l_{44} = \left( \frac{\Phi_0}{2\pi\mu_0 j_c} \right)^{1/2} B^{1/4}$, so that the grain size $d < l_{44}/2$, flux lines have less constraints relative to their alignment along the pinning centers and, hence, have more freedom for a better matching with the point defects. Therefore, from the lowest field $H > H_{c1}$ the vortices are very well pinned, the degree of disorder is increased as regarding the bulk Bragg glass, and any field induced softening of the shear modulus by increasing the field strength does not bring a significant energetic gain as occurs in the case of bulk Bragg glass-disordered vortex glass transition.

## 4. CONCLUSION

Intra-grain critical current density in $(Cu, C)Ba_2Ca_3Cu_4O_{10+\delta}$ was improved by the addition of 0.1 mol LiF and an appropriate increase in the oxygenation level. The effect is maximized in the $LiF_{0.1}$-added single-phase sample obtained by a careful matching of oxygen content at $z_{AgO} = 0.73$. The improvement has two features; *first*, a large second peak (FTE) develops on a considerable field range, and *second*, the limits of irreversibility are pushed to fields higher than in the LiF-free sample for $T > T_c/2$. When there is a deviation from the optimal oxygenation, the irreversibility is improved to a lesser level.

The fishtail effect depends on the deviation of the hole density from the (quasi) optimal doping, as defined by the LiF-*free* sample *P0*, and shifts to high fields when a certain level of homogenization is reached. The irreversibility field is also dependent on the homogenization of the distribution of the oxygen vacations. However, in both cases a certain level of disorder ($\gamma$) is necessary and its drastic reduction by over-oxygenation has adverse effects on both characteristics.

As expected, irreversibility effects are size dependent and the fishtail effect is significantly depressed in ground samples.

Although in this work we have studied the possibility to control and enhance the irreversibility characteristics for the Cu,C-1234 phase through additions such as LiF keeping the single-phase integrity of the samples, similar work would be interesting for the other phases from the $(Cu,C)Ba_2Ca_{n-1}Cu_nO_x$, $n \geq 1$ superconducting series. This is because each phase would probably have a different T-H domain to maximize irreversibility characteristics due to different level of disorder and doping. From a practical point of view it might be advantageous in some cases to produce and use in applications mixed samples with expanded T-H region of the maximized irreversibility characteristics through the overlapping of the irreversibility effects that are shown by the two or more component phases with additions.

**Acknowledgment**

Authors would like to thank Prof. K. Togano (IMR, Tohoku University, Japan) for reading the manuscript and for allowing the use of a SQUID magnetometer and to Dr. A. Crisan for continuous support of this work (NIMP, Romania). P.B. gratefully acknowledges STA/JSPS fellowship. This work was partially supported by Ministry of Education, Research and Youth, Romania under CERES program.

**Figure caption**

**Figure 1** Magnetization hysteresis loops for 20 K (solid) and 60 K (thick) in: a) - sample *P0*, (Cu,C) –1234, and b) - sample *P2*, (Cu,C) –1234- $(LiF)_{0.1}$.

**Figure 2** Paramagnetic susceptibility vs. AgO content ($z_{AgO}$) at 77K as obtained from equilibrium magnetization at high fields.

**Figure 3** Relative critical current density as obtained from Bean model at different temperatures and compositions: a)- sample *P0*, (Cu,C) -1234; b)- sample *P2* (Cu,C) –1234-$(LiF)_{0.1}$ single phase (The long arrow shows the decrease of the temperature. The short vertical arrows mark the peak field); c)- Critical current density for the samples *P0*, *P2*, *P3*, *P4*, *P5* and *P6* at 77 K. Inset: low field detail.

**Figure 4** Field dependence of the fishtail critical current density $J_{FT}$ at 77 K. Inset shows the amplitude $J_{FT,max}$ vs. $z_{AgO}$.

**Figure 5** Field dependence of the fishtail critical current density $J_{FT}$ for the optimally oxygenated LiF-doped (Cu,C)-1234 sample (*P2*) at temperatures between 40 and 102 K.

**Figure 6** Temperature dependence of the peak field for the optimally oxygenated LiF-doped (Cu,C)-1234 samples.

**Figure 7**. The irreversibility field vs. reduced temperature in the samples *P0* (LiF-free), *P2 and P3* (LiF-doped). Inset shows the irreversibility field vs. $z_{AgO}$ at 77 K.

**Figure 8** Magnetization width $\Delta M$ at 77 K for the sample *P0* (circles) and *P2* (triangles) vs. field in the as-prepared samples (filled symbols) and ground samples (empty symbols).

**Figure 9** Entry magnetization at 77 K of the sample *P0* (circles) and *P2* (triangles) vs. field in the as prepared samples (filled symbols) and ground samples (empty symbols).

**Table 1** Samples, starting (mol) LiF and AgO additions, phase composition of the samples, critical temperature and transition width (Cu,C-1234) as determined from the resistance vs. temperature curves and the room-temperature Seebeck coefficient.

| Sample | $y_{LiF}/z_{AgO}$ | Phase composition | $T_c$ (K) | $\Delta T_c$ (K) | $S_{293K}$ ($\mu VK^{-1}$) |
|---|---|---|---|---|---|
| *P0* | 0.0/0.45 | (Cu,C)-1234, | 116.3 | 0.48 | -3.6±1.3 |
| *P1* | 0.05/0.59 | (Cu,C)-1234 | 114.6 | 0.61 | -5.0±1.5 |
| *P2* | 0.10/0.73 | (Cu,C)-1234, $Ca_{0.8}CuO_2$ (traces) | 113.0 | 0.92 | -8.0±1.5 |
| *P3* | 0.15/0.73 | (Cu,C)-1234 | 111.2 | 0.60 | -7.4±2 |
| *P4* | 0.10/0.45 | (Cu,C)-1245, (Cu,C)-1234 (traces) | Two steps transition $T_c^{Cu,C-1234}$= 114.4K ($T_c^{midpoint}$=112K) | 1.04 | -0.3±1.0 |
| *P5* | 0.10/0.59 | (Cu,C)-1245, (Cu,C)-1234 | 114.3 | 1.08 | 0±2 |
| *P6* | 0.10/0.80 | (Cu,C)-1234, (Cu,C)-1223 (traces) | 113.0 | 0.95 | -4.3±1.0 |

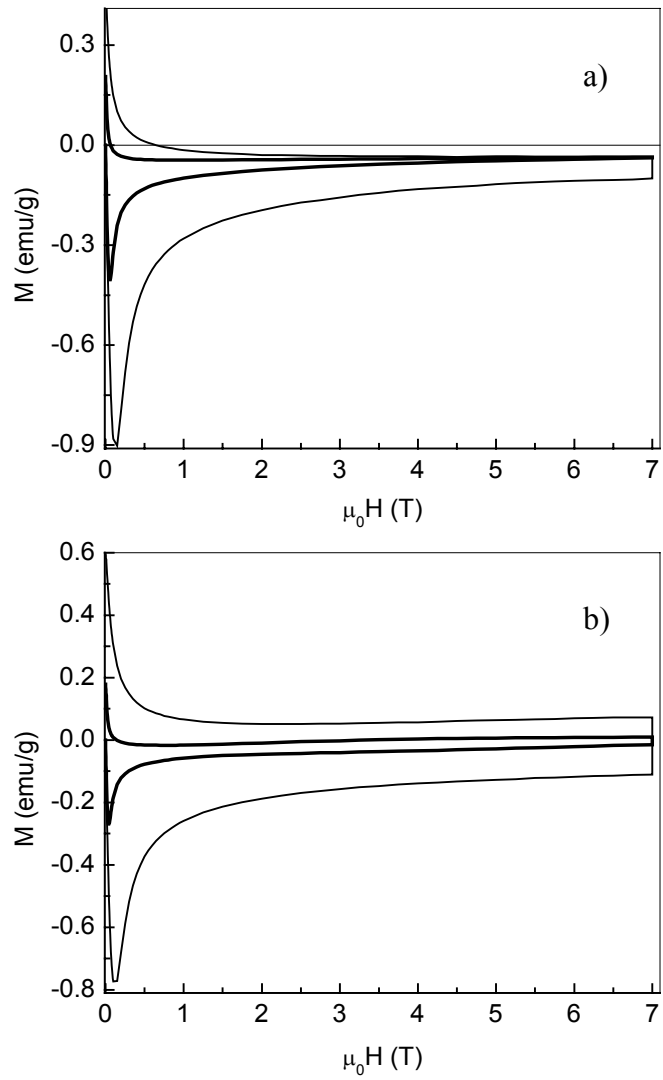

Fig. 1 P. Badica et al, Fish-tail effect and irreversibility field of…

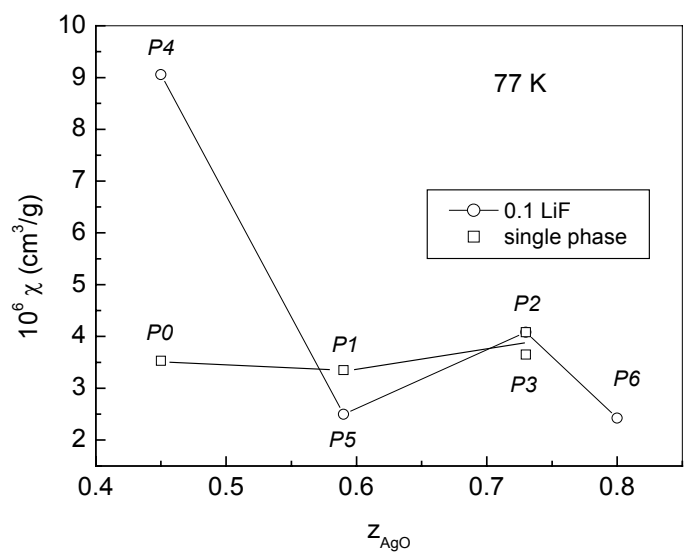

Fig. 2 P. Badica et al, Fish-tail effect and irreversibility field of…

\

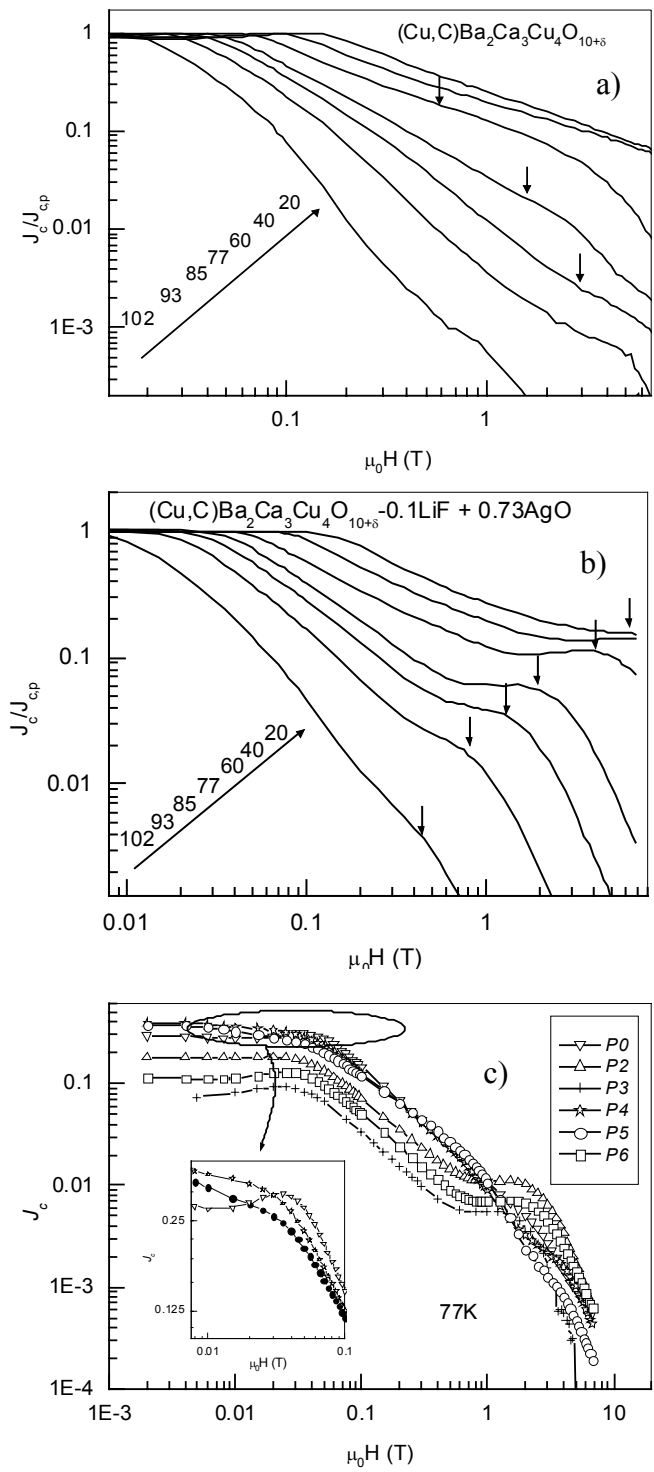

Fig. 3 P. Badica et al, Fish-tail effect and irreversibility field of…

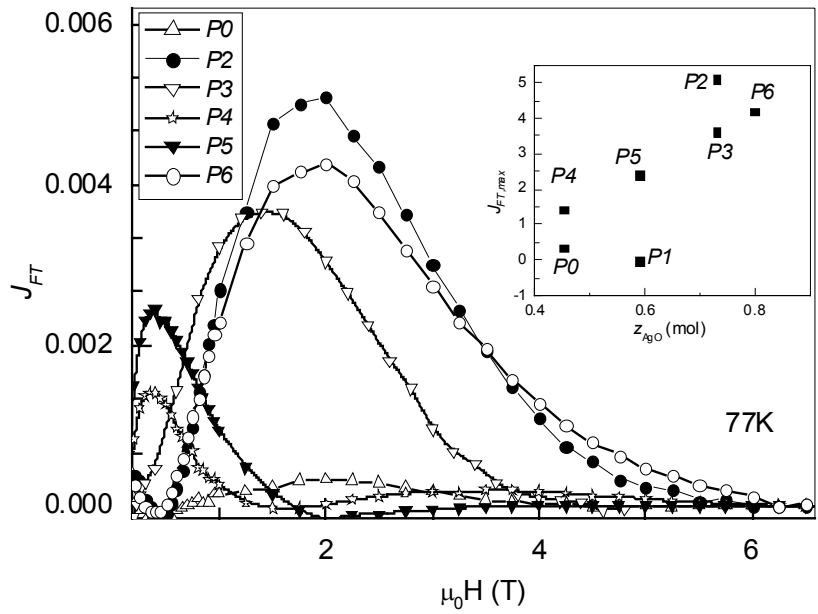

Fig. 4 P. Badica et al, Fish-tail effect and irreversibility field of…

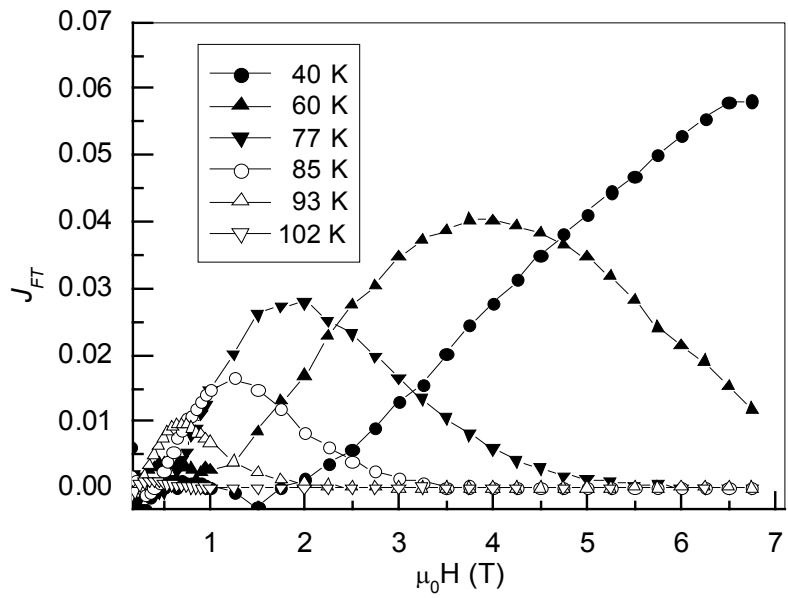

Fig. 5 P. Badica et al, Fish-tail effect and irreversibility field of…

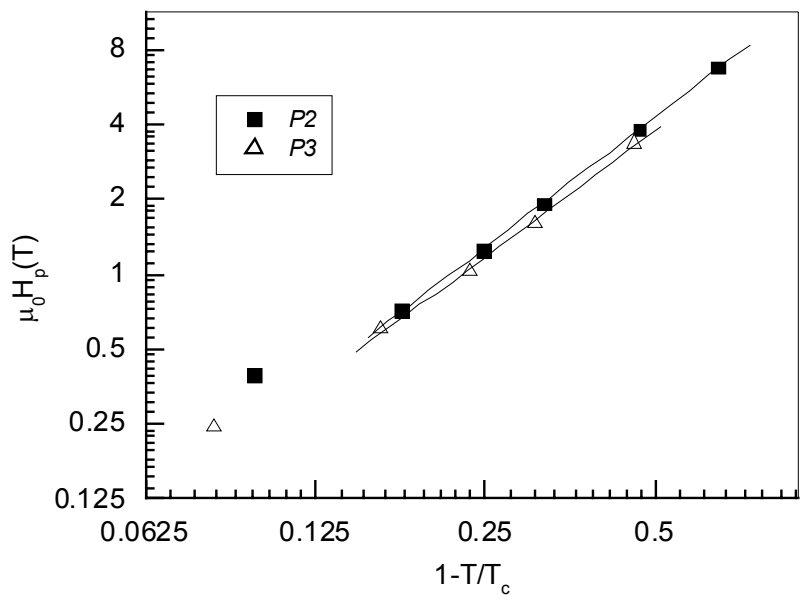

Fig. 6 P. Badica et al, Fish-tail effect and irreversibility field of…

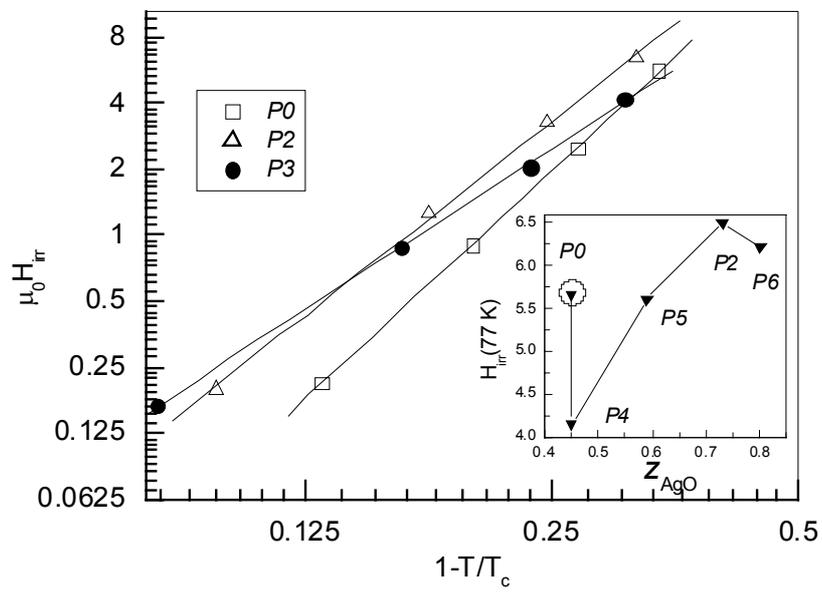

Fig. 7 P. Badica et al, Fish-tail effect and irreversibility field of…

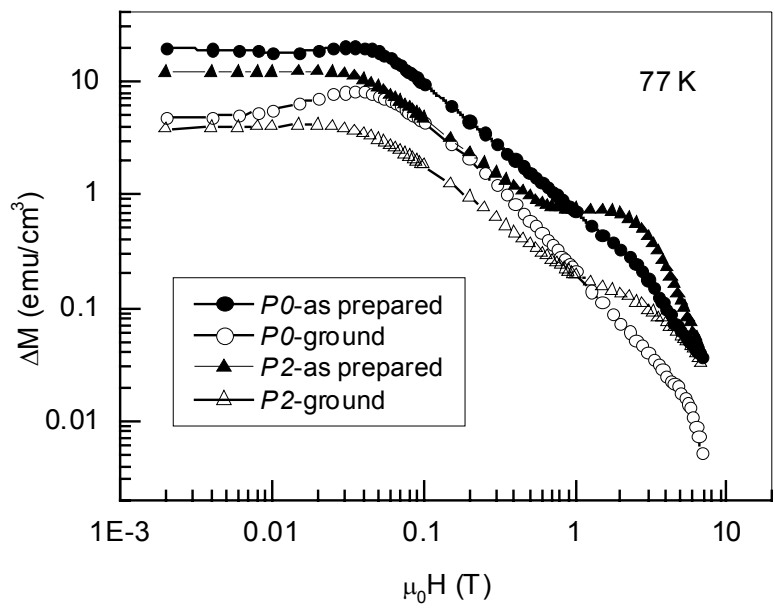

Fig. 8 P. Badica et al, Fish-tail effect and irreversibility field of…

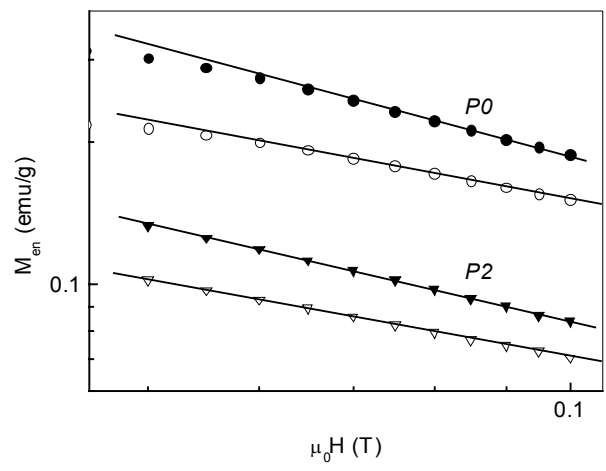

Fig. 9  P. Badica et al, Fish-tail effect and irreversibility field of…